\documentclass{article}

\usepackage{arxiv}

\usepackage[utf8]{inputenc} 
\usepackage[T1]{fontenc}    
\usepackage{hyperref}       
\usepackage{url}            
\usepackage{booktabs}       
\usepackage{amsfonts}       
\usepackage{nicefrac}       
\usepackage{microtype}      
\usepackage{lipsum}		
\usepackage{graphicx}
\usepackage{natbib}
\usepackage{doi}

\usepackage{amsmath}
\usepackage{float}
\usepackage{subcaption}
\usepackage{wrapfig}
\graphicspath{ {./figures/} }
\usepackage{enumerate}
\usepackage{diagbox}
\usepackage{natbib}
\usepackage{hyperref}
\usepackage{array}
\usepackage{url}

\newcommand{\cm}[1]{\ignorespaces}

\def\bfI{\mathbf I}

\def\bfL{\mathbf L}

\def\bfN{\mathbf N}

\def\bfalpha{\boldsymbol \alpha}
\def\bfbeta{\boldsymbol \beta}

\def\bfOmega{\boldsymbol\Omega}

\title{Bayesian learning of COVID-19 Vaccine safety while incorporating Adverse Events ontology}

\author{ Bangyao Zhao \\
	Department of Biostatistics, University of Michigan\\
	\And
Yuan Zhong \\
Department of Biostatistics, University of Michigan\\
	\And
Jian Kang \thanks{Corresponding author}\hspace{.2cm}\\
Department of Biostatistics, University of Michigan\\
	\And
Lili Zhao \thanks{Corresponding author}\hspace{.2cm}\\
Department of Biostatistics, University of Michigan\\
}

\renewcommand{\shorttitle}{Bayesian learning of COVID-19 Vaccine safety}

\hypersetup{
pdftitle={A template for the arxiv style},
pdfsubject={q-bio.NC, q-bio.QM},
pdfauthor={David S.~Hippocampus, Elias D.~Striatum},
pdfkeywords={First keyword, Second keyword, More},
}

\begin{document}
\maketitle

\begin{abstract}
While vaccines are crucial to end the COVID-19 pandemic, public confidence in vaccine safety has always been vulnerable. Many statistical methods have been applied to VAERS (Vaccine Adverse Event Reporting System) database to study the safety of COVID-19 vaccines. However, all these methods ignored the adverse event (AE) ontology. AEs are naturally related; for example, events of retching, dysphagia, and reflux are all related to an abnormal digestive system. Explicitly bringing AE relationships into the model can aid in the detection of true AE signals amid the noise while reducing false positives. We propose a  Bayesian graphical model to estimate all AEs while incorporating the AE ontology simultaneously. We proposed strategies to construct conjugate forms leading to an efficient Gibbs sampler. Built upon the posterior distributions, we proposed a negative control approach to mitigate reporting bias and an enrichment approach to detect AE groups of concern. The proposed methods were evaluated using simulation studies and were further illustrated on studying the safety of COVID-19 vaccines. The proposed methods were implemented in R package \textit{BGrass} and source code are available at \url{https://github.com/BangyaoZhao/BGrass}.
\end{abstract}


\section{Introduction}
\label{sec:intro}



As of February 1st, 2022, coronavirus disease 2019 (COVID-19) has infected over 376 million people and caused over 5.67 million deaths world-wide. The pandemic caused both high public attention and serious public health risk, and vaccines are crucial to end the pandemic. The US Food and Drug Administration (FDA) approved two mRNA-based COVID-19 vaccines, the BNT162b2 (Pfizer–BioNTech) and the mRNA-1273 (Moderna) and the virus-based COVID-19 vaccine Ad26.COV2.S (Johnson \& Johnson–Janssen) is still under emergency use authorization (EUA).






Despite the high safety standard of the FDA on COVID-19 vaccines, the public confidence in vaccines is always fragile. Due to the limited sample size, restrictive inclusion criteria, and limited duration of follow-up in the phase-3 trials,  rare and serious outcomes associated with COVID-19 vaccines may not be identified. Besides, although theoretically proven to be safe, the two mRNA-based vaccines (BNT162b2 and mRNA-1273) are brought to the market for the first time, and their safety can only be testified by long-term public surveillance.

Vaccine Adverse Event Reporting System (VAERS), accepting spontaneous reports of suspected vaccine adverse events after administration of any vaccine licensed in the United States from 1990 to present, is an important data source to study vaccine safety.  VAERS reports were received primarily from vaccine manufacturers, state and local health departments, and health-care providers, with fewer reports filed directly by patients and parents or others and are publicly available on the VAERS website \url{www.vaers.hhs.gov/data/index}.

VAERS is a key component in providing early warnings of vaccine safety problems. This large database (748,779 reports for the year 2021 alone) provides a valuable resource to detect safety problems (called ``signals") related to COVID-19 vaccines. A signal is considered evidence that an adverse event (AE) might be caused by vaccination and warrants further investigation or action. Several AE signals have been identified after COVID-19 vaccines using VAERS data, including anaphylaxis and myocarditis after receipt of mRNA-based COVID-19 vaccines \citep{shimabukuro2021reports, oster2022myocarditis}, and Guillain-Barré syndrome and Cerebral Venous Sinus Thrombosis (CVST)  after receipt of the Ad26.COV2.S vaccine \citep{shimabukuro2021reports,woo2021association}. 

As of December 24, 2021, the VAERS database included a total of 710295 AE case reports for all FDA-authorized or approved COVID-19 vaccines, including 319082 reports for the Pfizer COVID-19 vaccine, 326403 for the Moderna COVID-19 vaccine, 63224 for the Janssen COVID-19 vaccine, and 1586 for COVID-19 vaccines for which the manufacturer is unknown. VAERS data is commonly described by a large contingency table. In this table, each row represents a vaccine and each column represents an AE. Each cell in the table contains the number of reports that mention both that vaccine and that AE for a pre-defined period. Such a table can be very large, with hundreds of rows for vaccines and thousands of columns for AEs. Since we are interested in studying safety of the COVID-19 vaccines, the large table can be collapsed into a $2\times 2$ table for each AE with rows corresponding to COVID-19 vaccines (Yes/No) and two columns for the AE (Yes/No). The two rows can also be COVID-19 vs influenza vaccines, or mRNA-based vaccines vs Janssen COVID-19 vaccine.  The strength of the vaccine-AE association is commonly expressed in terms of a ratio, such as relative risk or Odds Ratio (OR)  \citep{zhao2020improvement}. A large OR ($>1$) indicates a safety signal. For example, if the $OR=3$ when comparing COVID-19 versus influenza vaccines, then the odds of the AE  occurred after COVID-19 vaccines is three times more likely than the odds of AE after influenza vaccines.

Statistical methods used to test the vaccine-AE association include the chi-squared test, likelihood ratio test \citep{huang2011likelihood},  Gamma-Poisson Shrinker (GPS) model \citep{DuMouchel:1999}, Multi-item GPS model \citep{DuMouchel:2001}, and Bayesian confidence propagation neural network (BCPNN) \citep{Bate:1998,Orre:2000, Noren:2006}.  However, none of the existing methods considered the inherent connections among AEs.  AEs are naturally related. For example, events of retching, dysphagia and reflux are all related to an abnormal digestive system. Explicitly bringing AE relationships into the model will aid in the detection of true signals amid the noise while reducing false positives. 

Several public resources are available to be used as ontology both to describe AEs and to form the basis of a statistical model. The largest resource for AE ontology is MedDRA (Medical Dictionary for Regulatory Activities), which describes disease relationships using a multi-level hierarchy.  The second lowest term, ``Preferred Terms" (PT), is a distinct descriptor for a symptom, sign and disease. Related PTs are grouped into ``High Level Terms" (HLT), ``High Level Group Terms" (HLGT) and  the largest category ``System Organ Classes" (SOC). Higher layers represent biologically and clinically meaningful categories for the disease observed on the lower levels. Multiple AEs may individually be rare enough to be undetected, but if they are related, they can borrow strength from each other to increase the chance for each individual AE to be detected.  In this paper, we model data on the PT level and use the higher HLGT level  to build connections between AEs on the PT level. Specifically, we create an interconnected network by linking two AE terms on the PT level  if they belong to the same AE group on the HLGT level. This network naturally considers the complexity that each PT AE term might belong to multiple groups.

In this article, we use logistic regression to model the binary AE outcome as a function of the binary vaccine type and confounders and then obtain the adjusted OR. We may have hundreds or even thousands of adjusted ORs, and one for each AE. We propose to build connections between these adjusted ORs by converting the above AE relationship network into the prior distribution of these parameters. To perform simple and fast Bayesian inference, we used Pólya-gamma augmentation and re-parameterized the covariance to construct conjugate forms for efficient Gibbs sampling.  Moreover, we proposed a negative control approach to mitigate the reporting bias in VAERS data and propose an approach to identify enriched AE groups where signaled AEs in the group occur more frequently than expected. The proposed methods were implemented in R package \textit{BGrass} and source code are available at \url{https://github.com/BangyaoZhao/BGrass}.

The rest of the article is organized as follows. In section 2 we introduce notation and develop our proposed methods. In section 3 we present simulation studies. In section 4 we illustrate our methods in studying safety of COVID-19 vaccines using VAERS database. Concluding remarks are given in section 5.

\section{Methods}
\label{sec:meth}
\subsection{Notation and  Model}
Suppose there are $n$ reports in VAERS that mentioned the vaccines of interests. For example, the report mentioned COVID-19 (target) or influenza (control) vaccines if we are interested in comparing these two types of vaccines. Of these reports, a total of $J$ AEs were mentioned. Let $A_{ij}$ be the indicator of mentioning AE $j$ on report $i$, i.e., $A_{ij} = 1$ if AE $j$ is mentioned on report $i$ and 0 otherwise. We used a series of  logistic regression models (one for each AE) to assess the associations between AEs and vaccines:

\begin{align}
&\mbox{logit}(Pr(A_{ij}=1|V_i,\boldsymbol{X_i}))=
\boldsymbol{\alpha}_j^{\top}\mathbf{X}_i +\delta_j \beta^*_j V_{i}
\quad \mbox{for}\ i = 1,\ldots, n,\ j = 1,\ldots,J \label{eq::logistic}\\
&\hspace{3cm}(\beta^*_1,...,\beta_J^*)\sim\boldsymbol{N}(\boldsymbol{0},\boldsymbol{\boldsymbol{W^\top\Omega_\epsilon W}})\label{eq::raw beta prior}\\
&\hspace{3cm}\delta_j\sim Bern(\pi_j)
\quad \mbox{for}\ j = 1,\ldots, J, \label{eq::delta prior}\\
&\hspace{3cm}\alpha_{jl}\sim N(0,\sigma_{\alpha,l}^2)
\quad \mbox{for}\ j = 1,\ldots, J, \quad l = 0,\ldots, p, \label{eq::alpha prior 1} \\
&\hspace{4cm}\sigma_{\alpha,l}^{-2}\sim
IG(a_{\alpha},b_{\alpha})
\quad \mbox{for}\ l = 0,\ldots, p,\label{eq::alpha prior 2}
\end{align}
where $\mathbf{X}_i = (1,X_{i1},\ldots, X_{ip})^{\top} $, $\bfalpha_j = (\alpha_{j0},...,\alpha_{jp})$, $V_i$ is the indicator of vaccine type on report $i$.  The logOR for the $j^{th}$ AE is $\beta_j= \delta_j \beta_j^*,$ where $\delta_j$ is the indicator variable for selecting the $j^{th}$ AE as a safety signal ($\delta_j=1$ if logOR of $j^{th}$ AE is different from zero and $\delta_j=0$ otherwise). Equation (2)-(5) show prior distributions. $Bern(\pi)$ denotes the Bernoulli distribution with probability $\pi$, $N(0,\sigma_{\alpha,l}^2)$ denotes normal distribution with variance $\sigma_{\alpha,l}^2,$ and $IG(a_{\alpha},b_{\alpha})$ denotes the inverse gamma distribution with shape parameter, $a_{\alpha}$, and scale parameter, $b_{\alpha}$. 
 
We introduce correlation between logORs by assuming that $\boldsymbol{\beta}=(\beta_1,...\beta_J)^\top$ follows a multivariate normal distribution with means zero  and covariance matrix $\boldsymbol{W^\top\Omega_\epsilon W}$. This  covariance has the correlation component,  $\boldsymbol{\Omega_\epsilon},$ and variance component $W^\top W,$ where  $\boldsymbol{W}=diag(\sigma_{\beta,1},...,\sigma_{\beta,J})$ is the diagonal matrix with its $(j,j)$-th element being the standard deviation of $\beta_j^*.$ Correlation matrix, $\boldsymbol{\Omega_\epsilon},$ consists of the relation network described by the AE ontology (such as MedDRA) and a parameter, $\epsilon,$ controlling the degrees of information borrowing from the network.  This parametrization of the covariance matrix greatly facilitates the posterior sampling (see details in section \ref{samplew}).
 
This proposed model (1)-(5) allows us to directly incorporate AE ontology into estimation of logORs while selecting AE signals. We call this model ``Bayesian Graphical Model for Signal Selection" (BGrass). If we assign  independent normal priors for  $\beta_j^*$'s,  BGrass reduces to a collection of separate logistic regression models with the selection feature, abbreviated as Bss.

\textbf{Determination of $\Omega_\epsilon.$}
The AE relation network can be represented by a undirected graph $G(V,E)$, where $V$ is the set of vertices that correspond to the $J$ AEs, and $E=\{j\sim k\}$ is the set of edges where $j\sim k$ indicates AE $j$ and $k$ belong to the same AE group. Let $d_j$ represent the degree of vertex $j$, i.e. $d_j = \sum_{k = 1}^J I(j\sim k)$. Mathematically, we summarize the graph $G$ by a normalized graph Laplacian matrix $L$.
\begin{equation}
    L=
    \left\{ 
  \begin{array}{ c l }
    1 & \quad \textrm{if } j=k \\
    -1/\sqrt{d_jd_k}&\quad \textrm{if }j\sim k\\
    0                 & \quad \textrm{otherwise}
  \end{array}
\right.
    \label{eq:laplacian matrix}
\end{equation}

We then add a small positive number $\epsilon$ to the diagonal of $L$ to guarantee the positive definiteness (i.e., $\bfL+\epsilon \bfI_J$), which is commonly used as the precision matrix \citep{Chung:1997,li2008network,sun2010bayesian}. Finally, we convert this precision matrix to the correlation matrix $\Omega_\epsilon,$ and parameter $\epsilon$ controls the strength of information borrowing between similar AEs. When $\epsilon$ is small, the correlation matrix relies more on the AE relation network, $\bfL$, thereby, inducing a stronger degrees of information borrowing. If $\epsilon$ goes to $\infty$, BGrass reduces to Bss, where logORs of AEs are assumed to be independent. In this paper, we consider $\epsilon$ as a tuning parameter and  determine its value based the deviance information criterion (DIC). For example, we ran models with a wide range of $\epsilon$, say $\{10^{-3},10^{-2}, 10^{-1}, 10^{1},1,10,100, \infty\}$ and selected the model with the smallest DIC value. Once $\Omega_\epsilon$ is determined, we just need to estimate the vector of standard deviation parameters $(\sigma_{\beta,1},...,\sigma_{\beta,J})$ in $\boldsymbol{W}$.

\subsection{Posterior sampling}

\subsubsection{Construction of efficient Gibbs sampling for $\boldsymbol{W}$}
\label{samplew}

It is extremely challenging to  sampling $\boldsymbol{W}=diag(\sigma_{\beta,1},...,\sigma_{\beta,J})$ as $\Omega_\epsilon$ is not a diagonal matrix. To perform efficient sampling, we rewrite $\beta^*_j$  as $\beta_j^*=\beta_j^{**}\sigma_{\beta,j}$ in equation (\ref{eq::logistic}) and $\boldsymbol{\beta^{**}}\sim\boldsymbol{N}(\boldsymbol{0},\boldsymbol{\Omega_\epsilon}$) in  equation (\ref{eq::raw beta prior}),  such that equation (1) and (2) become
\begin{align*}
&\mbox{logit}(Pr(A_{ij}=1|V_i,\boldsymbol{X_i}))=\alpha_{j0} +
\boldsymbol{\alpha}_j^{\top}\mathbf{X}_i +\delta_j \beta_j^{**}\sigma_{\beta,j} V_{i}
\quad \mbox{for}\ i = 1,\ldots, n,\ j = 1,\ldots,J\\
&\hspace{3cm}(\beta^{**}_1,...,\beta_J^{**})\sim\boldsymbol{N}(\boldsymbol{0},\boldsymbol{\Omega_\epsilon})
\end{align*}

This reparametrization is mathematically equivalent to the original equation (1)-(2), but provides an efficient framework for posterior sampling. We assign $\sigma_{\beta_j}\sim FN(0,\tau^2)$ and $\tau^{-2}\sim IG(k/2,k/2)$, where $FN$ represents a folded normal distribution. This prior distribution is equivalent to a half-t prior distribution for $\sigma_{\beta,j}$ ($j = 1,...J$) with $k$ degrees of freedom  \citep{halfCauchy}.  Under this reparameterization,  we have simple conjugate forms to sample $\sigma_{\beta,j}$'s, which greatly simplifies the computation.

\subsubsection{Polya–Gamma data augmentation}

The likelihood function for the $j^{th}$ AE on report $i$ can be written as 
\begin{equation}
    \pi(A_{ij}|V_i,\boldsymbol{X_i})=\frac{exp\{A_{ij}(\boldsymbol{\alpha}_j^{\top}\mathbf{X}_i +  \delta_j\sigma_{\beta,j}\beta_j^{**}V_i)\}}{1+exp(\boldsymbol{\alpha}_j^{\top}\mathbf{X}_i + \delta_j\sigma_{\beta,j}\beta_j^{**}V_i)}
    \quad \mbox{for}\ i = 1,\ldots, n, \quad j = 1,\ldots, J,
    \label{eq::likelihood}
\end{equation}

This likelihood is analytically inconvenient and has long been considered as a challenging problem for Bayesian inference.  We apply a novel data-augmentation strategy proposed by \cite{polson2013bayesian} for posterior sampling; detailed definition of the Polya–Gamma distribution, denoted by $PG(b,c)$, can be found in \cite{polson2013bayesian}. Let $\psi_{ij}=\boldsymbol{\alpha_j^\top X_i}+V_i\delta_j\sigma_{\beta,j}\beta_j^{**}$ denote the linear function of predictors. A latent variable $\omega_{ij}\sim PG(1,0)$ is introduced for each $A_{ij}$ and  $(A_{ij},\omega_{ij})$ pairs are independently distributed with the following joint density function,
\begin{equation}
\pi(A_{ij},\omega_{ij}|V_i,\boldsymbol{X_i})\propto
exp\{(A_{ij}-\frac{1}{2})\psi_{ij}-\omega_{ij}\frac{\psi_{ij}^2}{2}\}f_{PG(1,0)}(\omega_{ij}),
\label{eq::augmented likelihood}
\end{equation}
where $f_{PG(1,0)}(\cdot)$ denotes the probability density function of a $PG(1,0)$-distributed random variable. Integrating out the $\omega_{ij}$ in the augmented likelihood function in (\ref{eq::augmented likelihood}) gives the likelihood function in  (\ref{eq::likelihood}). The equation (\ref{eq::augmented likelihood}) provides exponential function forms with respect to $\boldsymbol{\alpha_j}$ and $\beta_j^{**}$, thus achieving the conjugacy. The auxiliary random variable $\omega_{ij}$'s also has conjugate forms,  $(\omega_{ij}|A_{ij})\sim PG(1,\psi_{ij})$.

Through reparametrization of the covariance matrix and data augmentation for the binomial likelihood, all  posterior sampling have conjugate forms, leading to a simple and efficient Gibbs sampler. The full posterior distributions can be found in Supplementary document B.

\subsection{Signal detection based on Posterior distributions}\label{sec::pos_inf}

In our model, logOR parameter $\beta_j$ represents strength of the signal for AE $j$, and the posterior probability $\delta_j$ indicates the likelihood of $\beta_j$ being different from zero (i.e. statistical significance). Therefore, we use both $\beta_j$ and $\delta_j$ to determine if AE $j$ is a safety signal. In general, we identify AE $j$ as a safety signal for the target vaccine if the posterior mean of $\beta_j$ ($\hat{\beta_j}$) is large  (say $\hat{\beta_j} >log(2)$) and the posterior selection probability, $\hat{\delta_j},$ is large. To determine the threshold for $\hat{\delta_j},$ we applied the FDR control methods \citep{fdr} to $\hat{\delta_j}$'s to restrict the expected Bayesian FDR at a pre-specified level $\alpha$.

\noindent\textbf{Negative control approach.} Like all spontaneous reporting database, VAERS data also has limitations. One of the limitations is reporting bias. Therefore, some AEs can be falsely recognized as safety signals as they are more likely to be reported compared to other vaccines. For example, COVID-19 vaccines get more public attention compared to influenza vaccines, therefore, AEs after COVID-19 vaccination are more likely to be reported than AEs after influenza vaccines. In such situations, negative control methods can be a powerful tool to reduce reporting bias. \citep{shixu,schuemie2014interpreting}. Negative controls (NCs) are AEs that are known not to be causally related to any vaccine based on clinical knowledge. The list of NCs create an empirical null distribution of logOR estimates representing no vaccine-AE association.  We expect the logOR of an AE signal is larger than the logOR estimates of NCs (i.e., taking an extremely large value in the null distribution). Assuming that the null distribution is normal and we quantify  extremity of an logOR if  $\beta_j>\mu_{ng}+2\sigma_{ng}$, where $\mu_{ng}$ and $\sigma_{ng}$ is the mean and the standard deviation of logOR estimates for NCs at each MCMC iteration. The posterior probability, $p(\beta_j>\mu_{ng}+2\sigma_{ng}),$  is the NC-adjusted selection probability (NCprob) of AE $j$ being an signal. We can also use the FDR method described earlier to determine the threshold for AE signal selection. 

\noindent\textbf{Identifying AE groups of concern.} In addition to identifying individual AE signals, we can also identify AE groups of concern based on the posterior distributions. Motivated by gene enrichment methods in the omic-analysis \citep{Subramanian15545, 10.3389/fgene.2020.00654}, we define an AE group $G$ as enriched if it contains more signaled AEs than the remaining AE groups. To this end, we create a $2\times 2$ table at each MCMC interaction with two rows for AEs to be signaled and unsignaled  and two columns for AEs to be in and not in group G. We then calculate the logOR for this table, denoted by $\gamma_G$. A large positive $\gamma_G$ indicates that group G is likely enriched. Averaged over all MCMC iterations, the posterior probability, $p(\gamma_G>log(2)),$ is used to identify if group G is enriched.  

\section{Simulations}
\label{sec:verify}

\subsection{Simulation I}

We first set up a small simulation study to illustrate our key idea on the information borrowing based on the graphical prior. Specifically, we simulated 70 AEs, 30 of them in group 1, 15 of them in group 2, and the remaining 25 AEs were not in any group (named as isolated AEs). We set logORs to be $1$, $-1$, and $0$ for AEs in group 1, group 2, and isolated AEs, respectively. We simulated 50 datasets and applied the BGrass and Bss models to each simulated dataset.  

In both BGrass and Bss models, the shape and scale parameter, $a_\alpha$ and $b_\alpha$, in the $IG$ distribution in equation  \ref{eq::alpha prior 2} were both set to $0.5$. The degrees of freedom was set to be one (i.e.,$k=1$) in the half-t prior distributions for $\sigma_{\beta,j}$'s.  The probability in Bernoulli prior in  \ref{eq::delta prior} was set to $\pi = 0.5$ for all AEs. In the BGrass model, we used DIC criteria to select the best $\epsilon$ in $\{10^{-3},10^{-2}, 10^{-1}, 10^{1},1,10,100, \infty\}$ (representing high to low degrees of information borrowing).  Gelman-Rubin diagnostic test was used to check the model convergence. 

\begin{figure}[h]
    \centering
    \includegraphics{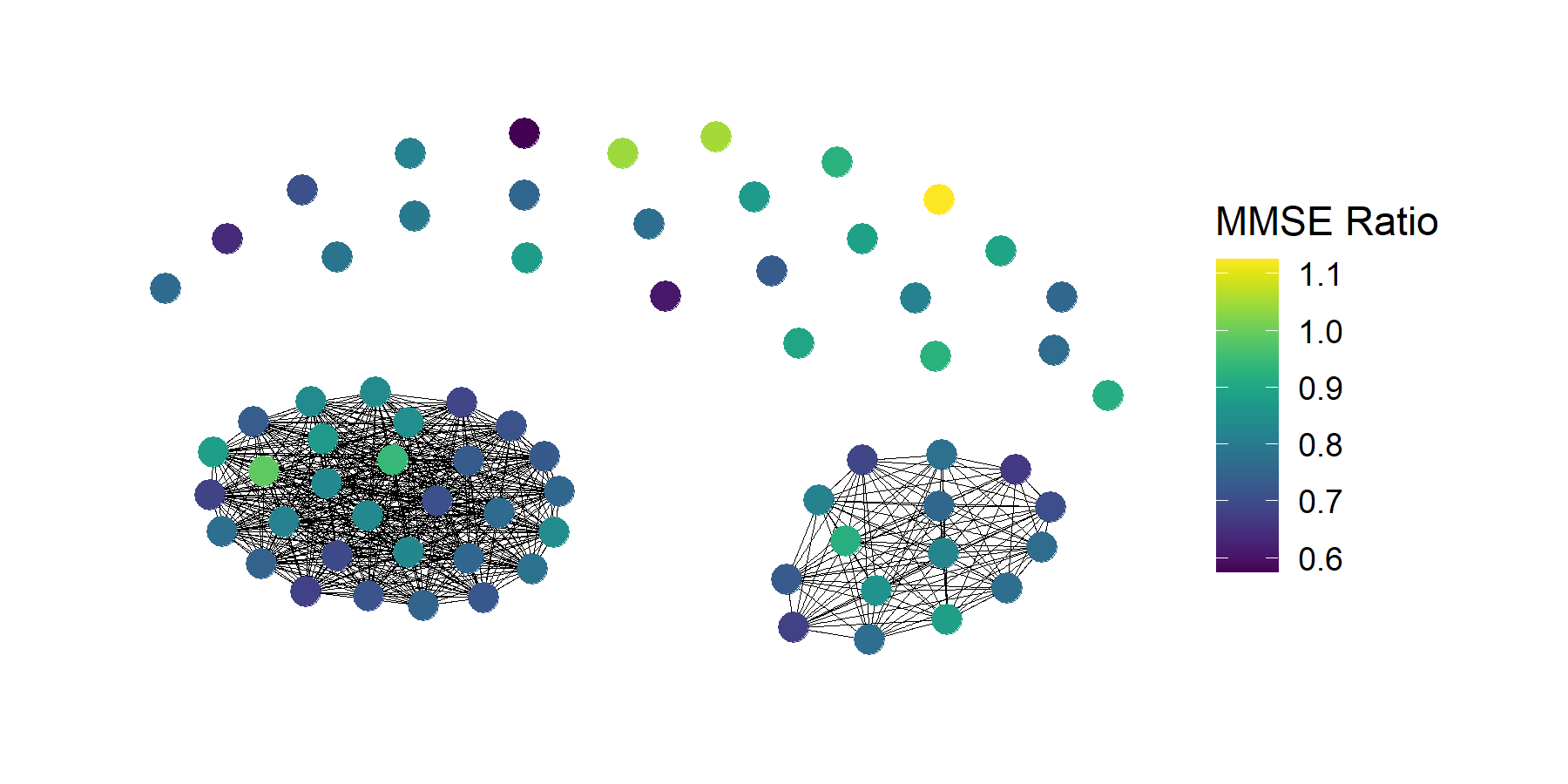}
    \caption{MMSE ratios (Brass versus Bss) for 70 AEs}
    \label{fig:sim1}
\end{figure}

We compared  our BGrass model with the Bss model (not including correlation between AEs)  using the averaged mean square error (MMSE). The MMSE for the $j$th AE was calculated by averaging the squared difference between the posterior mean $\hat{\beta_j}$ and true $\beta_j$ over the 50 simulated datasets. We visualized the MMSE ratios (BGrass over Bss) for all 70 AEs in Figure \ref{fig:sim1}. If the MMSE ratio for a particular AE is smaller than one, the BGrass model provides more accurate logOR estimate than the Bss model for that AE.  As Figure \ref{fig:sim1} shows, our BGrass model provides more accurate logOR estimates for vast majority of the AEs in group 1 and 2, indicating that the information borrowing between similar AEs help the parameter estimation. Surprisingly, isolated AEs also had more accurate logOR estimates in the BGrass model compared to the Bss model. One explanation is that the information borrowing improved the overall accuracy for the AE selection, thereby, increasing the accuracy for the logOR estimates.


\subsection{Simulation II}
\begin{figure}
    \centering
    \includegraphics{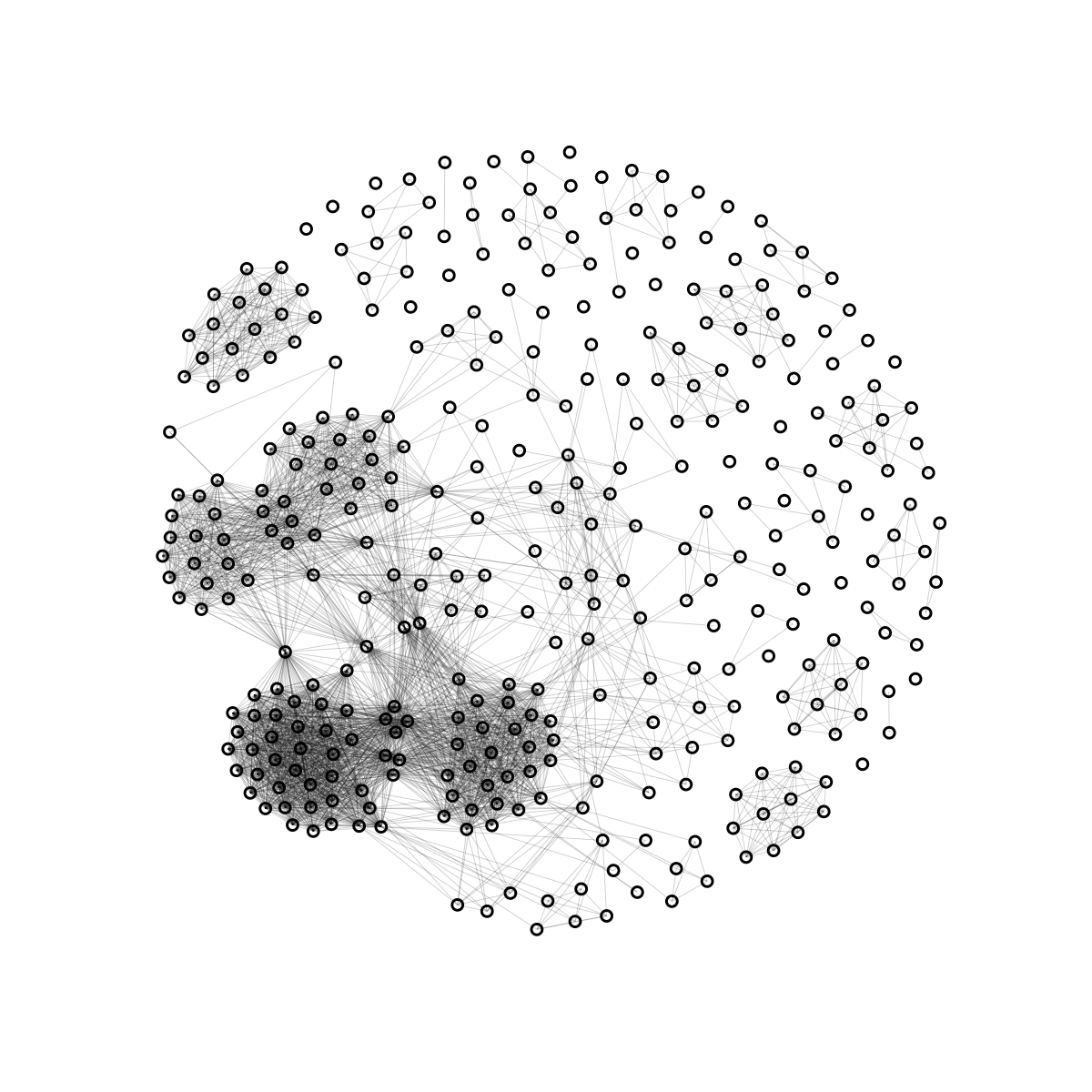}
    \caption{The relation network based on the HLGT level in MedDRA }
    \label{fig::sim2ontology}
\end{figure}

To mimic the real data, we randomly sampled 50,000 reports from the VAERS database to perform a larger simulation study. Those reports mentioned a total of 346 AEs, which were connected based on 78 AE terms on the higher HLGT level in MedDRA (the relation network is shown in Figure \ref{fig::sim2ontology}).

We generated data with different degrees of correlations between logOR. We first simulated $\bfbeta^*$ from $\bfN(\boldsymbol{0},0.1\bfOmega_\epsilon)$ with $\epsilon = \infty$ (no correlation), $\epsilon = 0.02$ (small correlation), and $\epsilon = 0.001$ (large correlation). Then, we simulated $\delta_j \sim Bern(0.5)$, and defined the true logORs to be $\beta_j=\delta_j \beta^*_j$. As a result, around 50\% of AEs had $\beta_j = 0$, called noise AEs, and 25\% of AEs had $\beta_j > 0$, called safety signals for the target vaccine. Under this setup, the averaged correlation between two connected non-noise AEs ($\beta_j \neq 0$) was 0 (no correlation), 0.12 (week correlation), and 0.56 (strong correlation). The regression coefficients $\bfalpha_j$'s (intercept, gender, and age) were chosen to mimic the real data. We also conducted another set of simulations with 80 \% of noise AEs and around 10\% of AE signals associated with the target vaccine.

\begin{figure}[h]
\includegraphics[width=1\linewidth]{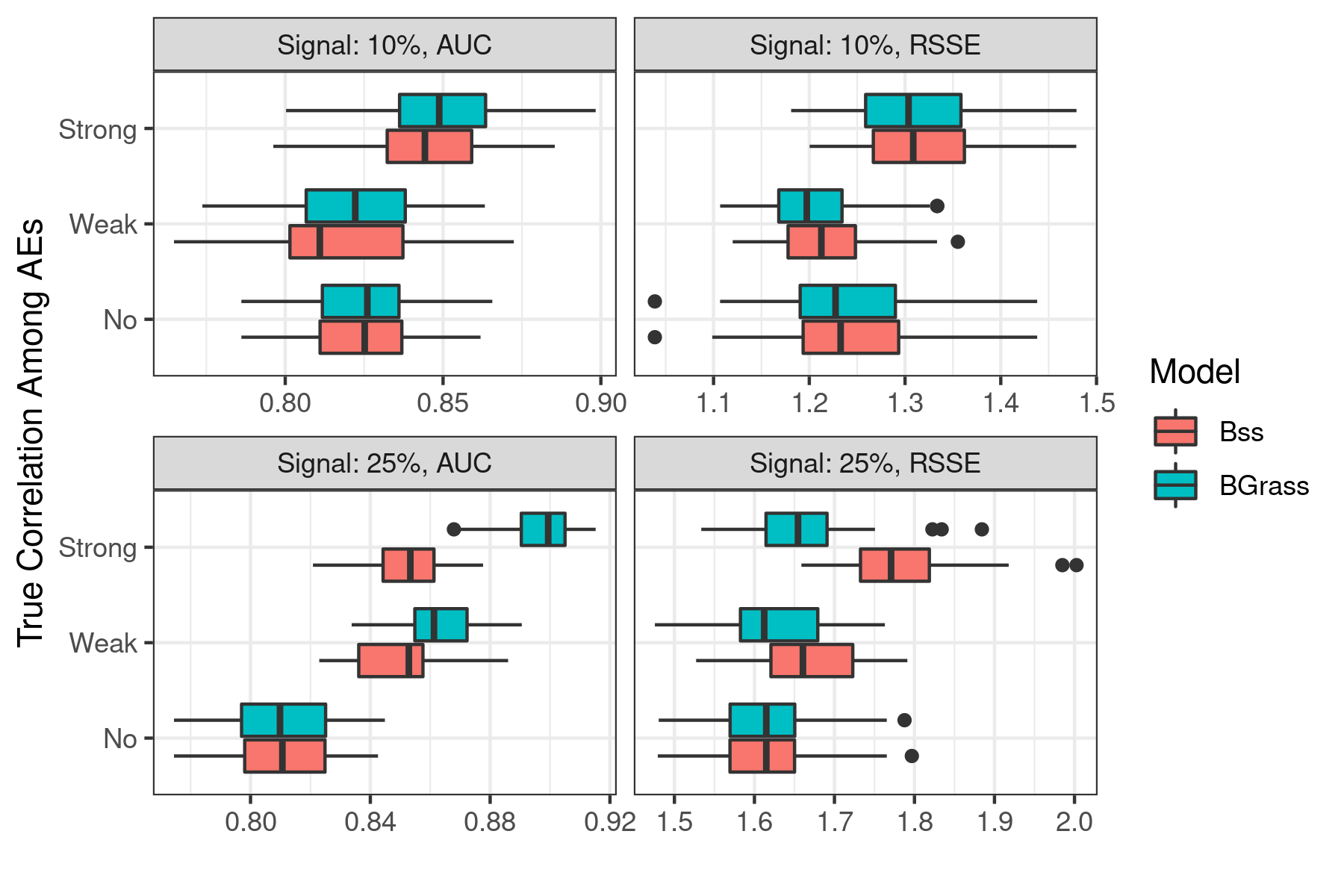}
\caption{RMSE and AUC Results from Simulation II}
\label{fig:sim2}
\end{figure}

We simulated 50 datasets for each scenario (3 correlation structures $\times$ 2 signal proportions). We applied both BGrass and Bss to each simulated data with hyperparameters the same as specified in the first simulation study. In each model, we ran three independent MCMC chains with 20,000 iterations and a burn-in of 10,000 and thinning of 2. We evaluated the model performance with regarding to the accuracy of the logOR estimates and the ability of correctly selecting AE signals. The accuracy was measured by the square root of sum squared error (RSSE),   $RSSE =\sqrt{\sum_{j=1}^J(\beta_j-\hat{\beta}_j)^2}$. For the signal selection ability, we computed posterior probability $P(\beta_j>0)$ for AE $j$ ($j=1,...,J$) and then used these values to construct the receiver operating characteristic (ROC) curve and  calculated the area under the curve (AUC). We presented the results of RMSE  and AUC in Figure \ref{fig:sim2}. This Figure demonstrates the clear advantage of incorporating the AE relations in the estimation. BGrass had improved performance compared to Bss, as evidenced by smaller RSSEs and larger AUCs, when AEs are correlated. As expected, they had similar performance when AEs are not correlated.  Moreover, BGrass's performance was further improved when  correlations are higher and signal proportions are larger. 

\section{Study safety of COVID-19 vaccines}

In the first study, we compared COVID-19 vaccines (Pfizer–BioNTech, Moderna and Johnson \& Johnson–Janssen) to  influenza vaccines (denoted as ``FLU", including inactivated influenza vaccines, labeled as``FLU3" and ``FLU4" and live attenuated influenza vaccine, labeled as``FLUN3" or ``FLUN4" in VAERS).  In the second study, we compared two mRNA-based COVID-19 vaccines (Pfizer–BioNTech and Moderna) to the virus-based COVID-19 vaccine (Johnson \&  Johnson–Janssen).  In all studies, we modelled AE data on the PT level and used the HLGT level of MedDRA to define AE groups. 

There are a total of 970,542 VAERS reports from January 1, 2016 to December 24, 2021. In each study, we removed reports with age under 18, mentioning both vaccines of interests (i.e,  a report with both COVID-19 and FLU vaccines was removed in the first study). We also removed AEs with a total report number less than 25.  We divided the age into four groups, [18,30), [30,50), [50,65), and $\geq$ 65.  With all categorical covariates, we counted the AE reports by gender, age group and vaccine type, aggregating a large number binary data into binomial data, which significantly improved the computation efficiency. 

We applied our BGrass model to both studies with the same hyperparameters as specified in the simulation studies. In BGrass, we ran 3 independent MCMC chains. For each chain, after a burn-in  of  10,000 iterations, we collected one MCMC sample in every 10 iterations until we got 20,000 MCMC samples. The MCMC chains in all studies have passed the Gelman-Rubin convergence diagnostic test with the maximum Gelman-Rubin statistic $R_c <1.1$. 

\subsection{COVID-19 v.s. FLU vaccines}

The final dataset contains 618,132 reports (585,760 for COVID-19 vaccines and 32,372 for FLU vaccines), and a total of 845 AEs, belonging to 118 AE groups. Gender distributions are similar between two vaccine groups and both groups was majority females (71.0\% in COVID-19  and  74.1\% in FLU). FLU vaccine recipients are older  (12.6\%, 34.6\%, 27.1\%, and 25.8\% for the pre-defined age groups in the COVID-19 group and  10.1\%, 22.4\%, 25.9\%, and 41.6\% in the FLU group). We set $\epsilon = 1$  based on the DIC criteria.

To mitigate the reporting bias due to the high public attention for the COVID-19 vaccines relative to the FLU vaccines, we used negative control approach to select AE signals. We identified $35$ AE terms in VAERS as negative controls (NCs) (see the list of NCs  in Supplementary document C). All of the NCs are known not to be causally related to vaccines and were reported more than 300 times in our dataset. The FDR for signal detection was controlled at  0.01. We also applied the  enrichment method to identify safety problem for AE groups containing more than 20 AEs, with FDR controlled at 0.01.

\begin{figure}[H]
    \centering
    \includegraphics[width=0.7\linewidth]{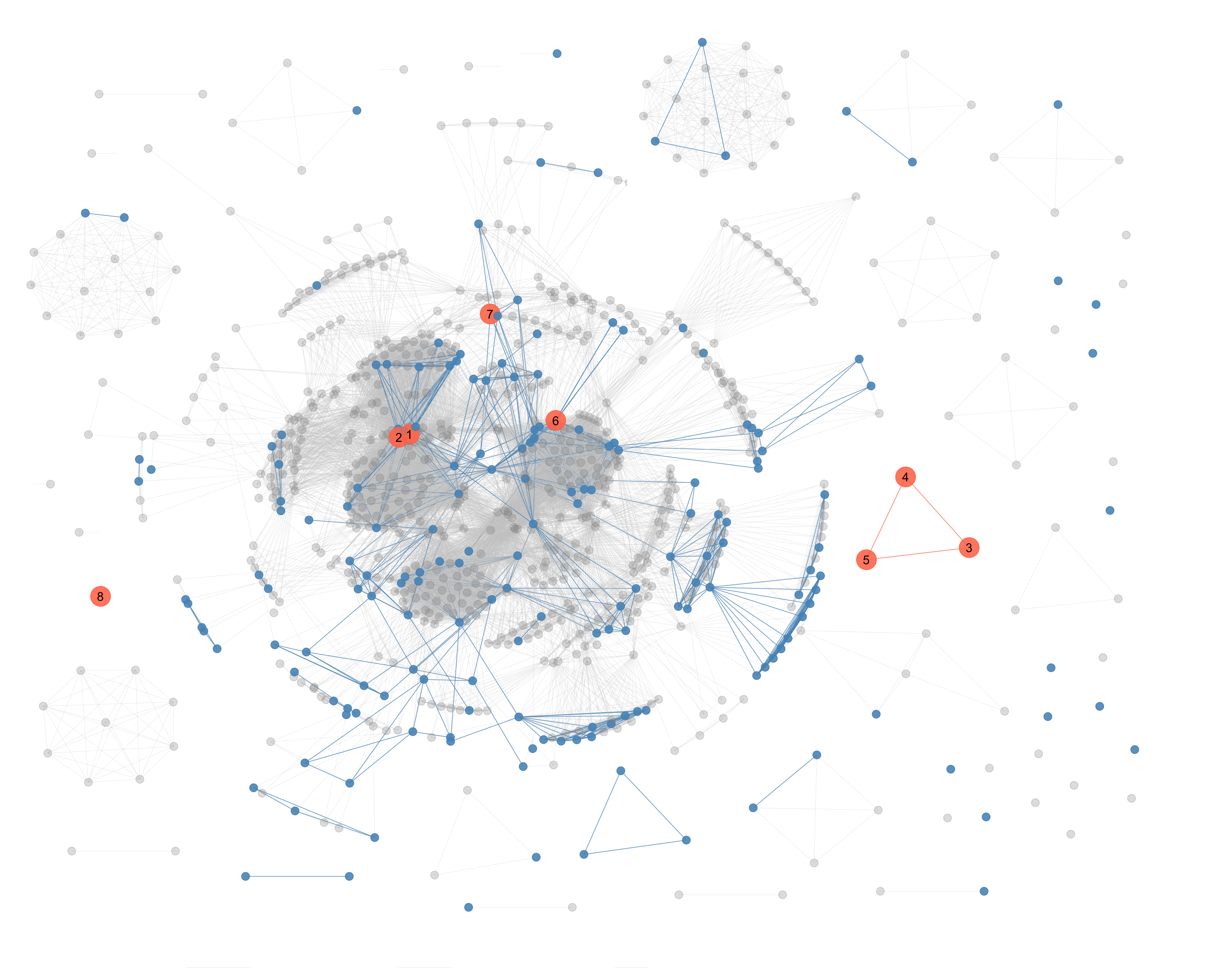}
    \caption{Study II (COVID-19 v.s. FLU): AEs with logOR $>log(2)$ (blue nodes) and signaled AEs identified using negative control (red nodes). The numbers on nodes map nodes to AEs in Table \ref{tb::study1} for reference. }
    \label{fig:study1network}
\end{figure}

We identified 8 AE signals in this study; see Table \ref{tb::study1} for the estimated logORs, NC-adjusted selection probability (NCprob). Figure \ref{fig:study1network} displays these AE signals in the relation network. Deep vein thrombosis, thrombosis and pulmonary embolism have been reported to be associated with COVID-19 vaccines \cite{see2021us, Hippisley-Coxn1931, 10.1001/jama.2021.21699}. Menstruation delay has also been reported to be associated with COVID-19 vaccines in a recent paper \cite{Menstruation}. Based on \cite{doi:10.1177/01455613211033125}, Ageusia (loss of sense of taste) and Anosmia (loss of sense of smell) are rare AEs associated with  COVID-19 vaccines. Since both of them are common symptoms of COVID-19 infection, they might be false positive signals. VAERS takes any adverse event that occurs after the administration of a vaccine, whether it is or is not caused by the vaccine. Therefore,  Ageusia and Anosmia might be related to the underlying disease rather than COVID-19 vaccines. Large epidemiological studies are needed to further study these signals. We didn't identify any enriched AE groups as there were only 8 safety signals using the negative control approach.

\begin{table}[H]
    \centering
    \begin{tabular}{cccc}\hline\hline
         & AE & logOR [95\% CI] & $NCprob$ \\\hline
        1 & Vaccination site pruritus &3.47 [3.27,3.66] &1 \\
        2 & Vaccination site erythema &3.16 [2.8,3.44]& 1\\
        3 &	Pulmonary embolism &3.49 [3.19,3.74]&1\\
        4 &Deep vein thrombosis &3.24 [2.64, 3.58]&0.995\\
        5 & Thrombosis & 2.58 [2.22, 2.96]&0.57\\
        6 & Menstruation disorder & 3.13 [2.43, 3.56]&0.876\\
        7 & Anosmia&2.92 [2.69, 3.16]&1\\
        8 & Ageusia&2.69 [2.4, 2.96]&0.669\\
\hline\hline
    \end{tabular}
    \caption{AE signals associated with COVID-19 vaccines compared to FLU vaccines. Visualization of these AEs in the relation network can be found in Figure \ref{fig:study1network}. }
    \label{tb::study1}
\end{table}

\subsection{mRNA-based Vaccines v.s. Johnson \& Johnson–Janssen}

The final dataset contains 588,487 reports (539,390 on mRNA-based vaccines and 49,097 for the Johnson \&  Johnson–Janssen vaccine), and a total of 845 AEs, belonging to 132 AE groups. There are majority females in both groups (71.9\% females in mRNA-based vaccine and 61.9\% in Johnson \& Johnson–Janssen). mRNA-based vaccine recipients are older than Johnson \& Johnson–Janssen recipients (12\%, 34.4\%, 26.8\%, and 26.8\% for the pre-defined age groups in mRNA-based vaccine recipients versus  18.8\%, 38.3\%, 30.2\%, and 12.7\% in Johnson \& Johnson–Janssen recipients). In this study, we are interested in AE signals associated with both mRNA-based vaccines and Johnson \& Johnson–Janssen  vaccines, represented by positive and negative logORs, respectively. We set $\epsilon =0.1$ based on the DIC criteria. 

\begin{figure}[H]
    \centering
    \includegraphics[width=0.7\linewidth]{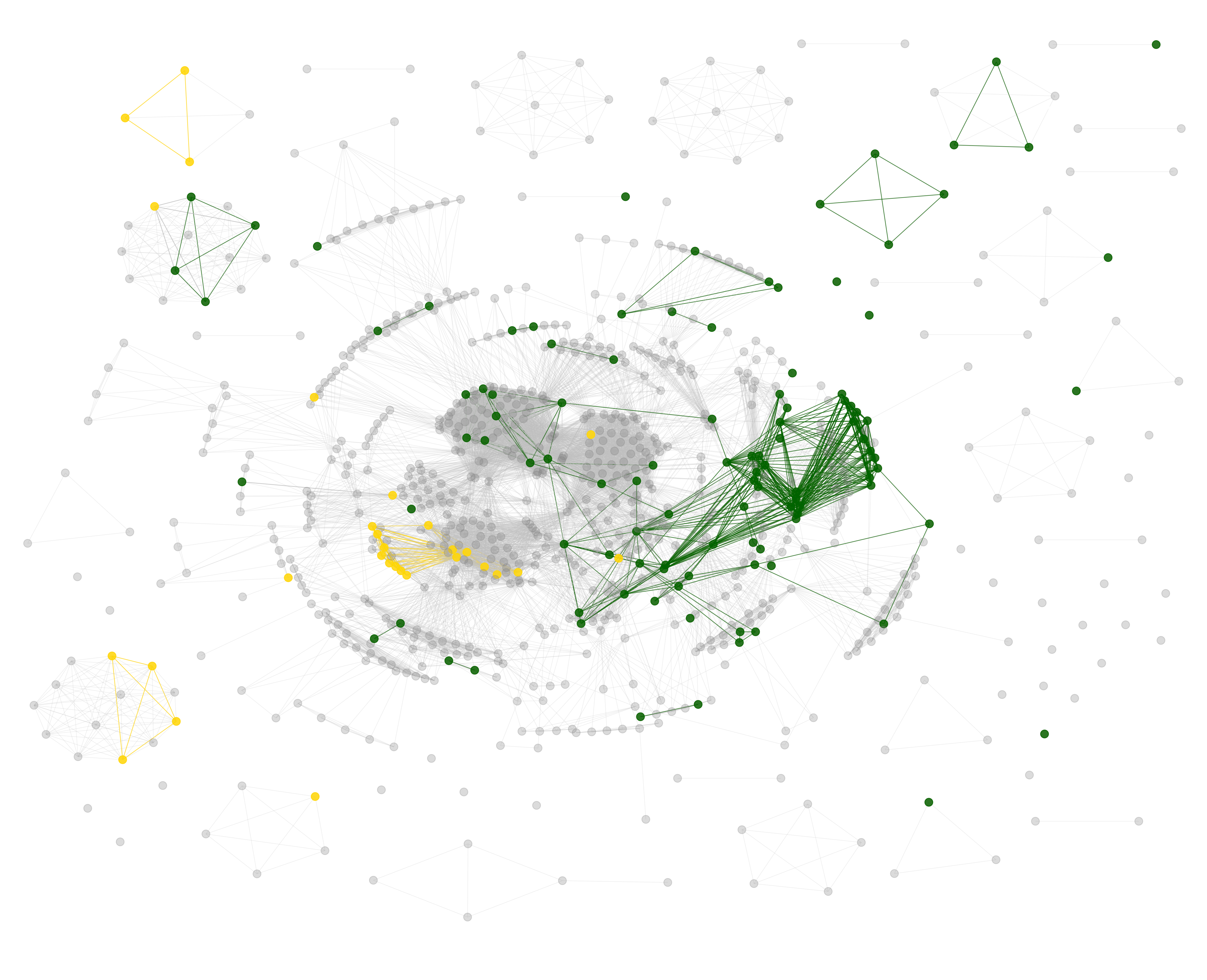} 
    \caption{Study II of comparing RNA Vaccines v.s. Johnson \&  Johnson–Janssen: yellow for AE signals associated with the mRNA-based vaccines and green for AE signals associated with the Johnson \& Johnson–Janssen vaccine }
    \label{fig:study2network}
\end{figure}

When FDR is controlled at 0.01, we identified 29 AEs associated with mRNA-based vaccines (represented by yellow in Figure \ref{fig:study2network}). Of them,  for example, thrombosis, myocarditis, cellulitis, rash erythematous, and throat irritation have been reported to be associated with mRNA-based vaccines \cite{doi:10.1161/CIRCULATIONAHA.121.056135,Ramalingam648,rash,KADALI2021376}.  We also identified 108 AEs associated with Johnson \& Johnson–Janssen (represented by green in Figure \ref{fig:study2network}). Of them, for example,  Guillain-Barre syndrome have been reported to be associated with Johnson \& Johnson–Janssen vaccines as mentioned in \href{https://www.fda.gov/news-events/press-announcements/coronavirus-COVID-19-update-july-13-2021}{CDC warning} and Cerebral venous sinus thrombosis (CVST) caused pause of the Johnson \& Johnson–Janssen vaccine \citep{shimabukuro2021reports,woo2021association}.  

For mRNA-based vaccines, we identified one enriched AE group ``administration site reactions". This group mainly contains mild and common symptoms, including 11 AE signals, such as injection site swelling, pain, and rash. For the Johnson \& Johnson–Janssen vaccine, we identified two enriched AE group, ``embolism and thrombosis" (including 20 AE signals) and ``central nervous system vascular disorders" (including 16 AE signals). see Supplementary material D for the complete list of AE signals.

In this study, we also applied the Gamma Poisson Shrinker (GPS) model in \cite{DuMouchel:1999} using the R package openEBGM. BGrass and GPS are not directly comparable, since they estimate different parameters (logOR in BGrass and Relative Risk in GPS). GPS also have no clear way to select AE signals. Therefore, we used both the relative rates and confidence intervals to select the AE signals; see details and comparsion results with BGrass in Supplementary material E.

\section{Conclusion}
\label{sec:conc}

Like all spontaneous reporting database, VAERS data also has limitations, such as incompleteness, inaccuracy, coincidence, difficulty in verification, and reporting bias. Nevertheless, due to its national scope, VAERS still serves as an important early warning system of vaccine safety issues. 

In this paper, we developed a graphical Bayesian model for detecting adverse events of COVID-19 vaccines while incorporating the AE ontology. The proposed BGrass model simultaneously estimates the strength of all AEs while allowing information borrowing between similar AEs. We developed novel methods to construct conjugate forms leading to an efficient Gibbs sampler for posterior computation. Simulation studies have demonstrated that BGrass has better performance compared to the model without considering the correlation between AEs. The analysis of VAERS data also identified important AE signals.

This article focus on COVID-19 vaccines, but the method is broadly applicable to other vaccine safety studies.

\bigskip
\begin{center}
{\large\bf SUPPLEMENTARY MATERIAL}
\end{center}

\begin{description}

\item[A. R-package for Gibbs sampling] This R package has passed the R CMD check and contains a vignette document to demonstrate its reproducibility. This package can be directly installed from \url{https://github.com/BangyaoZhao/BGrass}.

\item[B. Mathematical Details of the Gibbs Sampler] The detailed formulas for building the Gibbs sampler. (gibbs.pdf)

\item[C. Negative controls] The full list of $35$ negtive control AEs used in real data analysis I. (neg\_control.pdf)

\item[D. Signaled AEs in study II] The full list of signals identified in real data analysis II. (Study2\_MP\_JJ.pdf)

\item[E. Comparison with GPS] The comparison of BGrass and GPS on real data studies. (comparison.pdf)

\end{description}

\bibliographystyle{unsrtnat}
\bibliography{template}  






\end{document}